\documentclass[12pt]{article}
\usepackage{amssymb,amsmath,epsfig}
\usepackage{graphicx}
\allowdisplaybreaks
\begin{document}
\title{\bf Study of Viable Compact Stellar Structures in Non-Riemannian Geometry}
\author{M. Zeeshan Gul \thanks{mzeeshangul.math@gmail.com}~,
M. Sharif \thanks {msharif.math@pu.edu.pk}
and Adeeba Arooj \thanks{aarooj933@gmail.com}\\
Department of Mathematics and Statistics, The University of Lahore,\\
1-KM Defence Road Lahore-54000, Pakistan.}
\date{}
\maketitle

\begin{abstract}
The main objective of this article is to study the viable compact
stellar structures in non-Riemannian geometry, i.e.,
$f(\mathbb{Q},T)$ theory, where $\mathbb{Q}$ defines the
non-metricity and $T$ represents trace of the stress-energy tensor.
In this perspective, we consider a static spherical metric with
anisotropic matter configuration to examine the geometry of
considered compact stars. A specific model of this theory is used to
derive the explicit expressions of energy density and pressure
components that govern the relationship between matter and geometry.
The unknown parameters are evaluated by using the continuity of
inner and outer spacetimes to examine the configuration of spherical
stellar structures. Physical parameters such as fluid
characteristics, energy constraints and equation of state parameters
are analyzed to examine the viability of the considered stellar
objects. Further, we use Tolman-Oppenheimer-Volkoff equation, sound
speed and adiabatic index methods to analyze the equilibrium state
and stability of the proposed stellar objects. The rigorous analysis
and satisfaction of necessary conditions lead to the conclusion that
the stellar objects studied in this framework are viable and stable.
\end{abstract}

\section{Literature Review}

The general theory of relativity, formulated by Einstein is a
fundamental concept in physics that transformed our comprehension of
gravity and the structure of spacetime. It is a cornerstone of
modern physics and has been tested through observations and
experiments. However, this theory is based on geometric structures
in Riemann's metric space. Weyl \cite{1} developed a more general
geometrical structure that goes beyond Riemannian space and provides
a comprehensive explanation of gravitational fields and matter. His
objective was to unify gravitational and electromagnetic forces, not
all fundamental forces. The Levi-Civita connection is an essential
concept in Riemann metric space, which is used to compare vectors
based on their length. Weyl introduced a new type of connection that
does not consider the size of vectors during parallel transport. To
address the absence of information about vector's length, Weyl
introduced an additional connection known as the \emph{length
connection}. The length connection does not focus on the direction
of vector transport but instead on fixing or gauging the conformal
factor. Non-Riemannian geometries extend Riemannian geometry for
more general descriptions of spacetime curvature. These geometries
include torsion (twisting or rotation) or non-metricity (deviation
from metric compatibility). Weyl's theory incorporates the notion of
non-metricity with non-zero covariant derivative of the metric
tensor \cite{3}.

The non-metricity is a mathematical concept that emerges in theories
involving non-Riemannian geometries, providing an alternative cosmic
model without dark energy. In non-Riemannian gravity models metric,
co-frame and full connection are considered as gauge potentials. The
corresponding field strengths are the non-metricity
$\mathbb{Q}_{ij}$, the torsion $\mathbb{T}$ and the curvature
$\mathbb{R}_{ij}$. Because of the lack of experimental results for
the non-metricity and torsion, the non-Riemannian gravity models are
studied theoretically. Classification of the spacetime and related
theories are given in Table \textbf{1}. Researchers are drawn to
explore non-Riemannian geometry, specifically $f(\mathbb{Q})$
theory, for various reasons such as its theoretical implications,
compatibility with observational data and its significance in
cosmological contexts \cite{1aaa}. Recent investigations into
$f(\mathbb{Q})$ gravity have revealed cosmic issues and
observational constraints can be employed to indicate deviations
from the $\Lambda CDM$ model \cite{2aaa}. Spherical symmetric
configurations in $f(\mathbb{Q})$ gravity have been analyzed in
\cite{3aaa}. Ambrosio \cite{4aaa} described perturbation corrections
to the Schwarzschild solution in the same theory. Ambrosio et al
\cite{5aaa} delved into the asymptotic behavior of
Schwarzschild-like solutions in $f(\mathbb{Q})$ theory. The
non-metricity scalar has been employed to detect the effects of
microscopic systems in \cite{6aaa}. The viable cosmological
solutions in symmetric teleparallel gravity through the Noether
symmetry technique have been explored in \cite{7aaa}. Barros et al
\cite{8aaa} analyzed the cosmic characteristics through redshift
space distortion data in non-metricity gravity. This modified theory
can elucidate the cosmic bounce scenario \cite{9aaa} and describes
dark energy features at large scales \cite{10aaa}. For further
details, we refer the readers to \cite{11aaa}-\cite{16aaa}.
\begin{table}\caption{Classification of spacetimes.}
\begin{center}
\begin{tabular}{|c|c|c|}
\hline Relations & Spacetimes & Physical Theories
\\
\hline $\mathbb{Q}_{ij}=0$, $\mathbb{T}=0$, $\mathbb{R}_{ij}=0$&
Minkowski& Special Relativity
\\
\hline $\mathbb{Q}_{ij}=0$, $\mathbb{T}=0$, $\mathbb{R}_{ij}\neq0$ &
Riemannian & General Relativity
\\
\hline $\mathbb{Q}_{ij}=0$, $\mathbb{T}\neq0$, $\mathbb{R}_{ij}=0$ &
Weitzenbock & Teleparallel Gravity
\\
\hline $\mathbb{Q}_{ij}\neq0$, $\mathbb{T}=0$, $\mathbb{R}_{ij}=0$ &
&Symmetric Teleparallel
\\
\hline $\mathbb{Q}_{ij}\neq0$, $\mathbb{T}=0$,
$\mathbb{R}_{ij}\neq0$ & Riemann-Weyl & Einstein-Weyl
\\
\hline $\mathbb{Q}_{ij}=0$, $\mathbb{T}\neq0$,
$\mathbb{R}_{ij}\neq0$ & Riemann-Cartan& Einstein-Cartan
\\
\hline $\mathbb{Q}_{ij}\neq0$, $\mathbb{T}\neq0$,
$\mathbb{R}_{ij}\neq0$ & Non-Riemannian& Einstein-Cartan-Weyl
\\
\hline
\end{tabular}
\end{center}
\end{table}

Adak \cite{25aaa} studied the symmetric teleparallel gravity model
in which only non-metricity is non-zero. They obtained a spherically
symmetric static solution to Einstein equation in symmetric
teleparallel gravity and discussed the singularities. Nester and Yo
\cite{21aaa} studied teleparallel geometry with zero curvature and
torsion while non-zero nonmetricity behaves as the gravitational
force. Adak and Sert \cite{22aaa} explored a gravity model that is
characterized by nonmetricity, discovering that the horizon becomes
singular in symmetric teleparallel gravity. Adak et al \cite{23aaa}
formulated a symmetric teleparallel gravity model incorporating the
Lagrangian in the non-metricity tensor, comprehensively analyzing
the variations applicable to gravitational formulations. They
derived a set of solutions encompassing Schwarzschild,
Schwarzschild-de Sitter and Reissner-Nordstrom solutions for
specific parametric values. The spherical symmetric configuration in
$f(\mathbb{Q})$ gravity was investigated in \cite{17aaa}. Maurya et
al \cite{19aaa} noted the significant impact of the nonmetricity
parameter and decoupling constant on the stability of compact stars
in $f(\mathbb{Q})$ gravity. Adak et al \cite{24aaa} delved into the
broader realm of teleparallel geometry using differential forms.
Their exploration encompassed the examination of specific instances
such as metric and symmetric teleparallelism. They provided insights
into the connections between formulations employing gauge fixings
and those without gauge fixing. Additionally, the researchers
introduced a technique for transforming Riemannian geometries into
teleparallel structures. Adeel et al \cite{24aaaa} studied physical
analysis of anisotropic compact stars with different consideration
in this gravity

The presence of ghosts is a significant concern in modified
gravitational theories involving the non-metricity scalar. Ghosts
are theoretical entities that possess negative kinetic energy,
leading to instability and inconsistency in the theory. In the
context of gravitational theories, a ghost is associated with a
scalar field that contributes negative energy density. In
$f(\mathbb{Q})$ theory, the introduction of additional degrees of
freedom may lead to the emergence of ghost fields. These ghost
fields can result in unphysical solutions and inconsistencies in the
theory. The existence of ghosts raises questions about the overall
stability and predictability of $f(\mathbb{Q})$ gravity, casting
doubt on its ability to provide a consistent and physically
meaningful description of gravitational interactions. Addressing and
resolving the issue of ghosts in $f(\mathbb{Q})$ gravity is a
crucial task for researchers working on alternative gravitational
theories. Efforts are underway to formulate and refine
$f(\mathbb{Q})$ models in a way that eliminates or mitigates the
presence of ghosts, ensuring the theoretical soundness and
observational compatibility of these theories. The challenge lies in
constructing $f(\mathbb{Q})$ models that not only deviate from
general relativity but also maintain internal consistency and avoid
the emergence of undesirable ghost fields.

The modified symmetric teleparallel theory is further extended by
incorporating the trace of stress-energy tensor in the functional
action, named as $f(\mathbb{Q},T)$ theory \cite{4}. The
modifications introduced by $f(\mathbb{Q}, T)$ gravity have an
impact on the internal structure of compact stars. This influences
changes in the relationship between pressure and density, variations
in stellar radii and mass profiles. The corresponding equations of
motion play a role in hydrostatic equilibrium and affect the
stability of the star. Deviations from the predictions of general
relativity may rise to unique mass-radius relations that can be
tested against observational data from X-ray binaries. Neutron stars
are prime sources of gravitational waves in binary systems. The
modifications introduced by $f(\mathbb{Q}, T)$ gravity can lead to
distinct gravitational wave signatures that may differ from those
predicted by general relativity. These gravitational wave
differences as compared to general relativity predictions could be
explored using future gravitational wave detectors. Furthermore, the
implications of $f(\mathbb{Q}, T)$ gravity may extend to other
properties like surface redshift, providing avenues for
distinguishing this gravity model from other theories in the context
of compact stars.

Arora et al \cite{5} analyzed cosmic acceleration without additional
forms of dark energy in this theory. Arora and Sahoo \cite{5a}
examined accelerated and decelerated cosmic eras through the
deceleration parameter in this theory. Xu et al \cite{5d}
investigated that this modified theory predicts a de Sitter-type
cosmic expansion and represents an alternative to dark energy.
Najera and Fajardo \cite{5b} found that $f(\mathbb{Q},T)$ gravity
constitutes an alternative to the standard model of cosmology
($\Lambda$CDM). Godani and Samanta \cite{5c} studied the cosmic
evolution through different cosmological parameters (Hubble
parameter, deceleration parameter, luminosity distance, energy
conditions) and concluded that extended symmetric teleparallel
gravity represents the current cosmic accelerated expansion. Agrawal
et al \cite{5e} showed that the matter bounce scenario is possible
in this gravity. Tayde et al \cite{5f} used two different models of
this theory to study the existence of viable wormhole geometry in
extended symmetric teleparallel theory. Pradhan et al \cite{5g}
studied physical properties to ensure that a stable gravastar model
exists in this modified theory.

Stars are essential components of galaxies and maintain equilibrium
when the inner force (gravity) and the outer force (pressure)
produced from nuclear fusion reactions counterbalance each other
effects. Once a star's nuclear fuel is consumed, the insufficient
pressure leads to the formation of new remnants named as compact
objects. Researchers analyzed the evolutionary stages and internal
attributes of these dense objects by exploring their composition and
structure in the context of astrophysics. Baade and Zwicky argued
that stars are formed as a result of supernova explosions \cite{6}.
Pulsars (highly magnetized rotating neutron stars) provide evidence
for the existence of neutron stars \cite{7}. Pulsars emit
electromagnetic radiation beams and are observed as regular pulses
due to the neutron star's rotation. Neutron stars offer insights
into the behavior of matter under extreme densities and the effects
of strong gravitational fields. Herrera and Santos \cite{7a} studied
the impact of anisotropy on the geometry of compact objects. Rahaman
et al \cite{8} used the equation of state parameters to analyze the
viable features of compact stars. Hossein et al \cite{8a} used a
Krori-Barua solution with a radially dependent cosmological constant
to examine the geometry of pulsars. Harko et al \cite{8cc}
investigated the viability of pulsars through energy bounds and
examined their stable states using sound speed. The work in
different consideration has been studied in \cite{008cc}

The investigation of the physical characteristics of compact stellar
objects has been a subject of significant interest in the context of
modified gravitational theories. Olmo \cite{1aa} used a polytropic
EoS to study the properties of spherical stars in $f(\mathbb{R})$
gravity. Arapoglu et al \cite{2aa} employed the perturbation
technique to explore the geometry of compact stars in the same
theory. Shamir and Ahmad \cite{3aa} studied the physical properties
and stability of celestial objects in $f(\mathbb{G},T)$ theory.
Maurya et al \cite{4aa} examined the viable compact spherical
solutions in the framework of $f(\mathbb{R},T)$ theory. Biswas et al
\cite{5aa} discussed strange quark stars admitting the Krori-Barua
solution in the same theoretical framework. Bhar et al \cite{6aa}
used the Tolman-Kuchowicz solution to investigate the viable
characteristics of 4U 1538-52 compact star in Einstein Gauss-Bonnet
gravity. Sharif and Ramzan \cite{7aa} explored the behavior of
various physical quantities and the stability of distinct compact
stars in $f(\mathbb{G})$ theory. Rej et al \cite{8aa} examined the
possible features of charged SAX J 1808.4-3658 compact star in
$f(\mathbb{R},T)$ theory. Dey et al \cite{9aa} employed the
Finch-Skea ansatz to study viable anisotropic stellar models in
$f(\mathbb{R}, T)$ theory. Ilyas et al \cite{10aa} analyzed various
physical behaviors of the relativistic charged spheres, including
density profile and pressure components. Kumar et al \cite{11aa}
considered the Buchdahl model to analyze the structure of neutron
stars in this theory.

Nashed and Capozziello \cite{9a} formulated a new interior solution
for static spherically symmetric stars in the context of
$f(\mathbb{R})$ gravity and found that the corresponding interior
solution gives a viable neutron star model. Shamir and Malik
\cite{9b} analyzed the stability of charge Bardeen compact stars in
the same theory. Lin and Zhai \cite{9e} studied the impact of
effective matter variables on compact stellar structures in
$f(\mathbb{Q})$ theory. Ilyas \cite{18b} found that viable strange
stars exist in $f(\mathbb{R}, \mathbb{G},T)$ modified theory as all
the required conditions are satisfied ($\mathbb{G}$ is the
Gauss-Bonnet invariant). Malik \cite{18bb} investigated the behavior
of various physical quantities and stability of distinct compact
stars in $f(\mathbb{R},\phi)$ theory ($\phi$ is the scalar field).
Malik et al \cite{18bbb} analyzed the behavior of Her X1, SAX J
1808.4-3658 and 4U 1820-30 stars in $f(\mathbb{R},\phi, \chi)$
theory. Sharif et al examined the collapsing phenomenon \cite{018b},
stable regions of Einstein cosmos \cite{018bb} and exact solutions
by Noether symmetry approach \cite{018bbb} in $f(\mathbb{R}, T^{2})$
theory. Ilyas et al \cite{018bbbb} analyzed the geometry of charged
spherically symmetric strange stars to analyze the viability of the
considered stellar models in the same theory. Athar et al \cite{18c}
studied the viable geometry of anisotropic compact sphere in
$f(\mathbb{R}, \mathbb{G})$ theory.

Noether's symmetry approach, decoupling technique and embedding
method can be used to examine the geometry of compact objects. Deb
et al \cite{14aa} studied the geometry of anisotropic strange stars
through the embedding approach in $f(\mathbb{R},T)$ theory. Mustafa
et al \cite{15aa} analyzed anisotropic fluid spheres admitting the
same approach in $f(\mathbb{G},T)$ gravity. Maurya and Ortiz
\cite{16aa} employed a decoupling technique to analyze the physical
characteristics of compact stellar objects in $f(\mathbb{R},T)$
theory. Shamir and Naz \cite{17aa} considered the Noether symmetry
approach to examine the stability of anisotropic stellar structures
in modified $f(\mathbb{G})$ gravity. We have further extended this
work in modified $f(\mathbb{R},T)$ \cite{18aa} and
$f(\mathbb{R},T^{2})$ \cite{19aa} theories and found that the
obtained solutions depict the viability of proposed Noether
symmetric scheme. Azmat and Zubair \cite{20aa} employed a
gravitational decoupling approach to study the geometry of PSR
J1614- 2230, PSR 1937+21 and SAXJ1808.4-3658 compact stars in this
theory. The charged spherical solutions through the embedding
approach in $f(\mathbb{G},T)$ gravity have been discussed in
\cite{21aa}.

This literature motivates us to explore the viable characteristics
of anisotropic stellar structures in the context of
$f(\mathbb{Q},T)$ gravity. The following format is used in the
paper. Section \textbf{2} contains the basic formulation of
non-Riemannian geometry. We consider a specific model of this theory
to formulate the explicit expressions of energy density and pressure
components in section \textbf{3}. Section \textbf{4} determines the
physical characteristics of the considered stars using various
physical quantities. Section \textbf{4} examines the equilibrium
state and stability of the stars under consideration. We compile our
results in section \textbf{5}.

\section{Basics of Non-Riemannian Geometry}

This section presents the fundamental framework of the modified
$f(\mathbb{Q},T)$ theory and derives the field equations through the
variational principle. Weyl introduced a generalization of
Riemannian geometry as a mathematical framework for describing
gravitation in general relativity. The direction of a parallel
transported vector along a loop changes, but its length does not in
the Riemannian geometry. Weyl proposed a modification where a vector
can change its direction and size during parallel transport around a
closed path. This modification involves a new vector field $(h^{i})$
that characterizes the geometric properties of Weyl geometry. The
fundamental fields in Weyl's space are the new vector field and the
metric tensor. The metric tensor determines the local structure of
spacetime, defining distances and angles, while the vector field is
introduced to account for the change in length during parallel
transport.

In a Weyl geometry, if a vector length is transported with an
infinitesimal path then its length changes as $\delta l= lh_{i}
\delta x^ {i}$ \cite{19}. This indicates that the variation in the
vector's length is proportional to the original length, the
connection coefficient and the displacement along the path. The
variation in the vector's length after it is transported in parallel
around a tiny closed loop with area in the Weyl space is given as
\begin{equation}\label{1}
\delta l= l(\nabla_{j}h_{i}-\nabla_{i}h_{j})\delta s^{ij}.
\end{equation}
This states that the variation in the vector's length is
proportional to the original length, the curvature of the Weyl
connection and the area enclosed by the loop. A local scaling length
of the form $\bar{l}=\phi(x)l$ changes the field $h_{i}$ to
$\bar{h}_{i}=h_{i}+ (\ln\phi),_{i}$, whereas the elements of metric
tensor are modified by the conformal transformations
$\bar{\mathrm{g}}_{ij}=\phi^{2}\mathrm{g}_{ij}$ and
$\bar{\mathrm{g}}^{ij}=\phi^{-2}\mathrm{g}^{ij}$, respectively
\cite{20}. A semi-metric connection is another important feature of
the Weyl geometry, defined as
\begin{equation}\label{2}
\bar{\Gamma}^{u}_{~ij}=\Gamma^{u}_{~ij}
+\mathrm{g}_{ij}h^{u}-\delta^{u}_{~i}h_{j}- \delta^{u}_{~j}h_{i},
\end{equation}
where $\Gamma^{u}_{~ij}$ denotes the Christoffel symbol. One can
construct a gauge covariant derivative based on the supposition that
$ \bar{\Gamma}^{u}_{~ij}$ is symmetric. The Weyl curvature tensor
using the covariant derivative can be expressed as
\begin{equation}\label{3}
\bar{\mathbb{C}}_{ijuk}=\bar{\mathbb{C}}
_{(ij)uk}+\bar{\mathbb{C}}_{[ij] uk},
\end{equation}
where
\begin{eqnarray}\nonumber
\bar{\mathbb{C}}_{[ij]uk}&=&\mathbb{C} _{ijuk}+2\nabla_{u}h_{[{i}}
\mathrm{g}_{j]k}+2\nabla_{k}h_{[{j}}\mathrm{g}_{i]u}
+2h_{u}h_{[{i}}\mathrm{g}_{j]k}+2h_{k}h_{[{j}}\mathrm{g}_{i]u}
\\\nonumber
&-&2h^{2}\mathrm{g}_{u[{i}}\mathrm{g}_{j]k},
\\\nonumber
\bar{\mathbb{C}}_{(ij)uk}&=& \frac{1}{2}(\bar{\mathbb{C}}_{ijuk}
+\bar{\mathbb{C}}_{jiuk}).
\end{eqnarray}
The Weyl curvature tensor after the first contraction yields
\begin{equation}\label{4}
\bar{\mathbb{C}}^{i}_{~j}=\bar{\mathbb{C}}
^{ui}_{~uj}=\mathbb{C}^{i}_{~j}
+2h^{i}h_{j}+3\nabla_{j}h^{i}-\nabla_{i}h^{j}
+g^{i}_{~j}(\nabla_{u}h^{u}-2h_{u}h^{u}).
\end{equation}
Finally, we obtain Weyl scalar as
\begin{equation}\label{5}
\bar{\mathbb{C}}=\bar{\mathbb{C}}^{u}_{~u}=
\mathbb{C}+6(\nabla_{i}h^{i}-h_{i}h^{i}).
\end{equation}

Weyl-Cartan spaces with torsion represent a more generalized
framework beyond Riemannian and Weyl geometry. In this geometric
structure, the length of a vector is defined by a symmetric metric
tensor and the law of parallel transport is determined by an
asymmetric connection as $d\varpi^{i} =
-\varpi^{u}{\Gamma}^{i}_{~uj}dx^{j}$ \cite{21}. The connection for
the Weyl-Cartan geometry is expressed as
\begin{equation}\label{6}
\tilde{\Gamma}^{u}_{~ij}={\Gamma}^{u}_{~ij}
+\mathbb{W}^{u}_{~i}+\mathbb{L}^{u}_{~ij},
\end{equation}
where $\mathbb{L}^{u}_{~ij}$ is the deformation tensor and
$\mathbb{W}^{u}_{~ij}$ is the contortion tensor, defined as
\begin{equation}\label{7}
\mathbb{W}^{u}_{~ij}=\tilde{\Gamma}^{u}_{~[ij]}
+\mathrm{g}^{uk}\mathrm{g}_{iv}
\tilde{\Gamma}^{v}_{~[jk]}+\mathrm{g}^{uk} \mathrm{g}_{jv}
\tilde{\Gamma}^{v}_{~[ik]}.
\end{equation}
The non-metricity yields the deformation tensor as
\begin{equation}\label{8}
\mathbb{L}^{u}_{~ij}=\frac{1}{2}\mathrm{g}^{uk} (\mathbb{Q}_{jik}
+\mathbb{Q}_{ijk}-\mathbb{Q}_{uij}),
\end{equation}
where
\begin{equation}\label{9}
\mathbb{Q}_{uij}=\nabla_{u} \mathrm{g}_{~ij}
=-\partial\mathrm{g}_{ij,u}+\mathrm{g}_{~jk}
\tilde{\Gamma}^{k}_{~iu} +\mathrm{g}_{~ki}\tilde{\Gamma}^{k}_{~ju},
\end{equation}
and $\tilde{\Gamma}^{u}_{~ij}$ is Weyl-Cartan connection. Equations
(\ref{2}) and (\ref{6}) indicate that the Weyl geometry, where the
non-metricity is defined as
$\mathbb{Q}_{uij}=-2\mathrm{g}_{ij}h_{u}$ is a special case of the
Weyl-Cartan geometry with zero torsion. Therefore, Eqs.(\ref{6}) and
(\ref{7}) turn out to be
\begin{eqnarray}\label{10}
\tilde{\Gamma}^{u}_{~ij}&=&{\Gamma}^{u}_{~ij} +\mathrm{g}_{ij}h^{u}
-\delta^{u}_{~i}h_{j}-\delta^{u}_{~j}h_{i} +\mathbb{W}^{u}_{~ij},
\\\label{11}
\mathbb{W}^{u}_{~ij}&=&\mathcal{T}^{u}
_{~ij}-\mathrm{g}^{~uk}\mathrm{g}_{~vi}
\mathcal{T}^{v}_{~kj}-\mathrm{g}^{~uk} \mathrm{g}_{~vj}
\mathcal{T}^{v}_{~ki},
\end{eqnarray}
with
\begin{equation}\label{12}
\mathcal{T}^{u}_{~ij}=\frac{1}{2} (\tilde{\Gamma}^{u}
_{~ij}-\tilde{\Gamma}^{u}_{~ji}).
\end{equation}
The Weyl-Cartan curvature tensor is defined as
\begin{equation}\label{13}
\tilde{\mathbb{C}}^{u}_{~ijk} =\tilde{\Gamma}^{u} _{~ik,j}
-\tilde{\Gamma}^{u}_{~ij,k}+\tilde{\Gamma} ^{u}_{~ik}
\tilde{\Gamma}^{v}_{~uj}-\tilde{\Gamma} ^{u}_{~ij}
\tilde{\Gamma}^{v}_{~uk}.
\end{equation}
The contraction of this equation yields Weyl-Cartan scalar in the
following form
\begin{eqnarray}\nonumber
\tilde{\mathbb{C}}&=&\tilde{\mathbb{C}}^{ij} _{~ij}
=\mathbb{C}+6\nabla_{j}h^{j}-4\nabla_{j} \mathcal{T}^{j}-6h_{j}h^{j}
+8h_{j}\mathcal{T}^{j}+\mathcal{T}^{iuj} \mathcal{T}_{iuj}
\\\label{14}
&+&2\mathcal{T}^{iuj}\mathcal{T}_{jui}-4\mathcal{T}^{j}\mathcal{T}_{j}.
\end{eqnarray}

The gravitational action can be reformulated by eliminating the
boundary terms in the Ricci scalar as \cite{22}
\begin{equation}\label{15}
\mathcal{I}=\frac{1}{2\kappa} \int
\mathrm{g}^{ij}(\Gamma^{u}_{~ki}\Gamma^{k}_{~uj}
-\Gamma^{u}_{~ku}\Gamma^{k}_{~ij})\sqrt{-\mathrm{g}} d^ {4}x.
\end{equation}
Based on the assumption that the connection is symmetric
$(\Gamma^{u}_{~ij}=-\mathbb{L}^{u}_{~ij})$, we have
\begin{equation}\label{17}
\mathcal{I}=\frac{1}{2\kappa} \int
-\mathrm{g}^{ij}(\mathbb{L}^{u}_{~ki} \mathbb{L}^{k}_{~uj} -
\mathbb{L}^{u}_{~ku} \mathbb{L}^{k}_{~ij}) \sqrt{-\mathrm{g}} d^
{4}x,
\end{equation}
where
\begin{equation}\label{18}
\mathbb{Q}\equiv-\mathrm{g}^{ij}(\mathbb{L}^{u}_{~ki}
\mathbb{L}^{k}_{~uj} -\mathbb{L}^{u}_{~ku}\mathbb{L}^{k}_{~ij}),
\end{equation}
with
\begin{equation}\label{19}
\mathbb{L}^{u}_{~ki}\equiv-\frac{1}{2} \mathrm{g}^{uv}
(\nabla_{i}\mathrm{g}_{kv}+\nabla_{k} \mathrm{g}_{vu}
-\nabla_{v}\mathrm{g}_{ki}).
\end{equation}
From Eq.(\ref{17}), one can obtain the gravitational action of
$f(\mathbb{Q})$ theory by replacing non-metricity scalar with an
arbitrary function as
\begin{equation}\label{20}
\mathcal{I}=\int \frac{\sqrt{-\mathrm{g}}}{2\kappa}
f(\mathbb{Q})d^{4}x.
\end{equation}
This is the action of symmetric teleparallel theory, which is a
theoretical framework that provides an alternative geometric
description of gravity.

Now, we extend this gravitational Lagrangian as
\begin{equation}\label{22}
\mathcal{I}=\frac{1}{2\kappa}\int f(\mathbb{Q},T)
\sqrt{-\mathrm{g}}d^{4}x+\int
\mathcal{L}_{m}\sqrt{-\mathrm{g}}d^{4}x.
\end{equation}
The non-metricity scalar is defined as
\begin{eqnarray}\label{23}
\mathbb{Q}_{u}\equiv \mathbb{Q}^{~~i}_{u~~i}, \quad
\tilde{\mathbb{Q}}_{u}\equiv \mathbb{Q}^{i}_{~~ui}.
\end{eqnarray}
The superpotential of this model is given by
\begin{equation}\label{24}
\mathbb{P}^{u}_{~ij}=-\frac{1}{2}\mathbb{L} ^{u}_{~ij}
+\frac{1}{4}(\mathbb{Q}^{u} -\tilde{\mathbb{Q}}^{u})\mathrm{g}_{ij}-
\frac{1}{4} \delta ^{u} _{(i \mathbb{Q}_{j})},
\end{equation}
and the relation for $\mathbb{Q}$ is
\begin{equation}\label{25}
\mathbb{Q}=-\mathbb{Q}_{uij}\mathbb{P}^{uij}=-\frac{1}{4}
(-\mathbb{Q}^{ujk} \mathbb{Q}_{ujk}+2\mathbb{Q}^{ujk}
\mathbb{Q}_{kuj} -2\mathbb{Q}^{k}\tilde{\mathbb{Q}}_{k}+\mathbb{Q}
^{k}\mathbb{Q}_{k}).
\end{equation}
The calculation of the above relation is shown in Appendix
\textbf{A}.

In modified gravitational theories such as the $f(\mathbb{Q},T)$
theory, the coincident gauge is a specific choice of coordinate
system that simplifies the mathematical representation of the
theory. In the context of $f(\mathbb{Q},T)$ gravity, the theory
introduces additional terms involving the non-metric field and the
trace of the energy-momentum tensor. The coincident gauge is
considered to simplify the equations of motion and make the analysis
more tractable. It is a coordinate system where certain components
of the metric and other fields are chosen in such a way that the
equations governing the theory become more manageable. The
additional terms of $f(\mathbb{Q},T)$ theory allow for modifications
to the gravitational dynamics beyond what is predicted by Einstein's
theory. Thus, we assume a coincident gauge in our analysis to
simplify the field equations. The variation of Eq.(\ref{22}) with
respect to metric tensor yields
\begin{eqnarray}\nonumber
\delta\mathcal{I}&=&\int \frac{1}{2\kappa} \delta [f(\mathbb{Q},T)
\sqrt{-\mathrm{g}}]d^ {4}x+\int \delta [\mathcal{L}_{m}
\sqrt{-\mathrm{g}}] d^ {4}x,
\\\nonumber
&=&\int \frac{1}{2\kappa}\big[-\frac{1}{2}f\mathrm{g}_{ij}
\sqrt{-\mathrm{g}} \delta \mathrm{g}^{ij} + f_{\mathbb{Q}}
\sqrt{-\mathrm{g}} \delta \mathbb{Q} + f_{T} \sqrt{-\mathrm{g}}
\delta T
\\\label{26}
&-&\kappa T_{ij} \sqrt{-\mathrm{g}} \delta \mathrm{g}^{ij}\big]d^
{4}x.
\end{eqnarray}
The explicit formulation of $\delta\mathbb{Q}$ is given in Appendix
\textbf{B}. Moreover, we define
\begin{eqnarray}\label{27}
T_{ij} \equiv \frac{-2}{\sqrt{-\mathrm{g}}} \frac{\delta
(\sqrt{-\mathrm{g}} \mathcal{L}_{m})}{\delta \mathrm{g}^{ij}}, \quad
\Theta_{ij} \equiv \mathrm{g}^{uk} \frac{\delta T_{uk}}{\delta
\mathrm{g}^{ij}},
\end{eqnarray}
which implies that $ \delta T= \delta (T_{ij}\mathrm{g}^{ij}) =
(T_{ij}+ \Theta_{ij})\delta \mathrm{g}^{ij}$. Thus, Eq.(\ref{26})
turns out to be
\begin{eqnarray}\nonumber
\delta\mathcal{I}&=&\int \frac{1}{2\kappa}\bigg[\frac{-1}{2}f
\mathrm{g}_{ij}\sqrt{-\mathrm{g}} \delta \mathrm{g}^{ij} +
f_{T}(T_{ij}+ \Theta_{ij})\sqrt{-\mathrm{g}} \delta \mathrm{g}^{ij}
\\\nonumber
&-&f_{\mathbb{Q}} \sqrt{-\mathrm{g}} (\mathbb{P}_{iuk}
\mathbb{Q}_{j}^{~~uk}- 2\mathbb{Q}^{uk} _{~~i} \mathbb{P}_{ukj})
\delta \mathrm{g}^{ij}+2f_{\mathbb{Q}} \sqrt{-\mathrm{g}}
\mathbb{P}_{uij} \nabla^{u} \delta \mathrm{g}^{ij}
\\\label{28}
&-&\kappa T_{ij}\sqrt{-\mathrm{g}} \delta \mathrm{g}^{ij}\bigg]d^
{4}x.
\end{eqnarray}
The resulting field equations after equating the variation of this
equation to zero are
\begin{eqnarray}\nonumber
T_{ij}&=& \frac{-2}{\sqrt{-\mathrm{\mathrm{g}}}} \nabla_{u}
(f_{\mathbb{Q}}\sqrt{-\mathrm{g}} \mathbb{P}^{u}_{~ij})- \frac{1}{2}
f \mathrm{g}_{ij} + f_{T} (T_{ij} + \Theta_{ij})
\\\label{29}
&-&f_{\mathbb{Q}} (\mathbb{P}_{iuk} \mathbb{Q}_{~~j}^{uk}
-2\mathbb{Q}^{uk}_{~~i} \mathbb{P}_{ukj}),
\end{eqnarray}
where $f_{T}=\frac{\partial f}{\partial T}$ and
$f_{\mathbb{Q}}=\frac{\partial f}{\partial \mathbb{Q}}$. This
represents the field equations in $f(\mathbb{Q},T)$ theory and
solution of these equations can provide insights into how gravity
behaves in this modified framework.

\section{Field Equations and Matching Conditions}

We consider inner region as
\begin{equation}\label{28a}
ds^{2}=dt^{2}e^{\lambda(r)}- dr^{2}e^{\xi(r)}-r^{2}d\Omega^{2},
\end{equation}
where $d\Omega^{2}=d \theta^{2}+d\phi^{2}\sin^{2}\theta$. We
consider anisotropic matter distribution as
\begin{equation}\label{29a}
T_{ij}=\mathcal{U}_{i}\mathcal{U}_{j}
\rho+P_{r}\mathcal{V}_{i}\mathcal{V}_{j}-P_{t} g_{ij}+
P_{t}(\mathcal{U}_{i}\mathcal{U}_{j}
-\mathcal{V}_{i}\mathcal{V}_{j}).
\end{equation}
In gravitational physics, matter-Lagrangian density is a fundamental
concept that determines the configuration of matter and its dynamics
in a given spacetime. We consider the matter-Lagrangian density for
anisotropic matter as $\mathcal{L}_{m}=-\frac{P_{r}+2P_{t}}{3}$
\cite{23}. The chosen form of the Lagrangian density allows us to
capture the anisotropy. The resulting equations of motion are
\begin{eqnarray}\nonumber
\rho&=&\frac{1}{2r^{2}e^{\xi}}\bigg[2r\mathbb{Q}'f_{\mathbb{Q}
\mathbb{Q}}(e^{\xi}-1)
+f_{\mathbb{Q}}\big((e^{\xi}-1)(2+r\lambda')+(e^{\xi}+1)r\xi' \big)
\\\label{33}
&+&fr^{2}e^{\xi}\bigg]-\frac{1}{3}f_{T}(3\rho+P_{r}+2P_{t}),
\\\nonumber
P_{r}&=&\frac{-1}{2r^{2}e^{\xi}}\bigg[2r\mathbb{Q}'f_{\mathbb{Q}
\mathbb{Q}}(e^{\xi}-1)
+f_{\mathbb{Q}}\big((e^{\xi}-1)(2+r\lambda'+r\xi')-2r\lambda'\big)
\\\label{34}
&+&fr^{2}e^{\xi}\bigg]+\frac{2}{3}f_{T}(P_{t}-P_{r}),
\\\nonumber
P_{t}&=&\frac{-1}{4re^{\xi}}\bigg[-2r\mathbb{Q}'\lambda'f_{\mathbb{Q}
\mathbb{Q}} +f_{\mathbb{Q}}\big(2\lambda'(e^{\xi}-2)-r\lambda'^{2}
+\xi'(2e^{\xi}+r\lambda')
\\\label{35}
&-&2r\lambda''\big)+2fre^{\xi}\bigg]+\frac{1}{3}f_{T} (P_{r}-P_{t}).
\end{eqnarray}
These field equations are in complex form and we cannot deduce any
result from it. So, we take $f(\mathbb{Q},T)$ model as \cite{24}
\begin{eqnarray}\label{36}
f(\mathbb{Q},T)=\sigma\mathbb{Q}+\varsigma T,
\end{eqnarray}
where $\sigma$ and $\varsigma$ are arbitrary constants. The
resulting field equations (\ref{33})-(\ref{35}) are
\begin{eqnarray}\nonumber
\rho&=&\frac{\sigma
e^{-\xi}}{12r^2(2\varsigma^{2}+\varsigma-1)}\bigg[
\varsigma(2r(-\xi'(r\lambda'+2)+2r\lambda''+\lambda'(r\lambda'+4))-4e^{\xi}
\\\label{37}
&+&4)+3\varsigma r(\lambda'(4-r\xi'+r\lambda')+2r\lambda'')+12
(\varsigma-1)(r\xi'+e^{\xi}-1)\bigg],
\\\nonumber
P_{r}&=&\frac{\sigma
e^{-\xi}}{12r^2(2\varsigma^{2}+\varsigma-1)}\bigg[
2\varsigma\big(r\xi'(r\lambda'+2)+2(e^{\xi}-1)-r(2r\lambda''+\lambda'(r\lambda'
\\\nonumber
&+&4))\big)+3\big(r\big(\varsigma \xi'(r \lambda'+4)-2\varsigma
r\lambda''-\lambda'(-4\varsigma+\varsigma r\lambda
'+4)\big)-4(\varsigma-1)
\\\label{38}
&\times&(e^{\xi}-1)\big)\bigg],
\\\nonumber
P_{t}&=&\frac{\sigma
e^{-\xi}}{12r^2(2\varsigma^{2}+\varsigma-1)}\bigg[
2\varsigma\big(r\xi'(r\lambda'+2)+2(e^{\xi}-1)-r(2r\lambda''+\lambda'
\\\nonumber&\times&(r\lambda'+4))\big)+3\big(r
\big(2(\varsigma-1)r\lambda''-((\varsigma-1)r\lambda'-2)(\xi'-\lambda')\big)
\\\label{39}
&+&4\varsigma(e^{\xi}-1)\big)\bigg].
\end{eqnarray}

We consider Tolman IV solution as \cite{24a}
\begin{eqnarray}\label{40}
e^{\lambda(r)}=a^{2}r^{2n},  \quad
e^{\xi(r)}=\frac{n}{1-n(\frac{r}{c})^{d}},
\end{eqnarray}
where $n=1+2b-b^{2}$ and $d=\frac{2(1+2b-b^{2})}{1+b}$. The unknown
constants $(a, b, c)$ can be found using the Darmois junction
conditions. By imposing these conditions, researchers can model the
behavior of matter in celestial objects, leading to a deeper
understanding of their physical properties. We consider the outer
geometry of compact stellar objects as
\begin{eqnarray}\label{45}
ds^{2}_{+}=\aleph dt^{2}-\aleph^{-1}d r^{2}-r^{2}d\Omega^{2},
\end{eqnarray}
where $\aleph=1-\frac{2M}{r}$. The continuations of the first and
second fundamental forms at the surface boundary $(r=\mathcal{R})$
gives
\begin{eqnarray}\label{46}
\mathrm{g}_{tt}&=&a^{2}\mathcal{R}^{2n}= 1-\frac{2M}{\mathcal{R}},
\\\label{47}
\mathrm{g}_{rr}&=&\frac{n}{1-n(\frac{\mathcal{R}}{c})^{d}}=
(1-\frac{2M}{\mathcal{R}})^{-1},
\\\label{48}
\mathrm{g}_{tt,r}&=&
2a^{2}n\mathcal{R}^{2n-1}=\frac{2M}{\mathcal{R}^{2}},
\\\label{48a}
P_{r}(r=\mathcal{R})&=&0.
\end{eqnarray}
By solving these equations, we obtain
\begin{eqnarray}\label{49}
a=\sqrt{\frac{M}{b\mathcal{R}^{2b+1}}}, \quad
b=\frac{M}{\mathcal{R}-2M}, \quad
c&=&\mathcal{R}\bigg[\frac{\mathcal{R}(\mathcal{R}^{2}
-2M\mathcal{R}-M^{2})}{M^{2}(\mathcal{R}-2M)}\bigg]^{\frac{1}{d}}.
\end{eqnarray}

The compatibility of the solution is ensured by the non-singular and
positively increasing the behavior of metric elements throughout the
domain. The observed values of mass and radius of the considered
stars are given in Table \textbf{2} and the constants are shown in
Table \textbf{3}. Figure \textbf{1} shows that the behavior of both
metric potentials is positively increasing as required. In all
graphs, we use magenta, orange, purple, cyan, yellow, pink, green,
brown, gray, black, blue, red  for 4U 1538-52, Her X-1, LMC X-4, 4 U
1820-30, Cen X-3, 4U 1608-52, PSR J1903+327, PSR J1614-2230, Vela
X-1, EXO 1785-248, SAX J1808.4-3658 and SMC X-4 compact stars,
respectively.
\begin{figure}
\epsfig{file=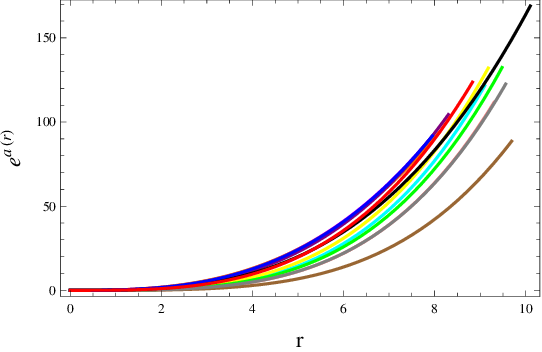,width=.5\linewidth}
\epsfig{file=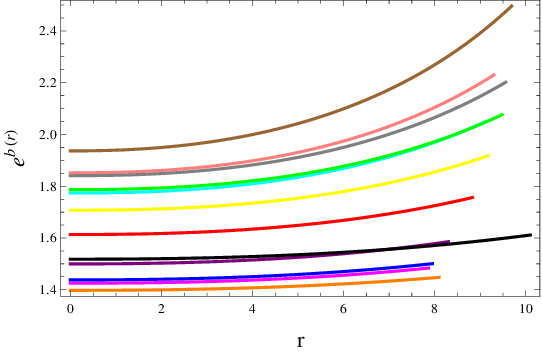,width=.5\linewidth}\caption{Plots of metric
elements for $\sigma=1.5$ and $\varsigma=-0.4$.}
\end{figure}
\begin{table}\caption{Values of input parameters.}
\begin{center}
\begin{tabular}{|c|c|c|c|c|c|c|c|}
\hline Compact stars & $M_{\odot}$ & $\mathcal{R}(km)$
\\
\hline 4U 1538-52 \cite{31} & 0.87 $\pm$ 0.07 & 7.866 $\pm$ 0.21
\\
\hline SAX J1808.4-3658\cite{32} & 0.9 $\pm$ 0.3 & 7.951 $\pm$ 1.0
\\
\hline Her X-1 \cite{33} & 0.85 $\pm$ 0.15 & 8.1 $\pm$ 0.41
\\
\hline LMC X-4 \cite{31} & 1.04 $\pm$ 0.09 & 8.301 $\pm$ 0.2
\\
\hline SMC X-4 \cite{31} & 1.29 $\pm$ 0.05 & 8.831 $\pm$ 0.09
\\
\hline 4U 1820-30 \cite{34} & 1.58 $\pm$ 0.06 & 9.1 $\pm$ 0.4
\\
\hline Cen X-3 \cite{31} & 1.49 $\pm$ 0.08 & 9.178 $\pm$ 0.13
\\
\hline 4U 1608-52 \cite{35} & 1.74 $\pm$ 0.01 & 9.3 $\pm$ 0.10
\\
\hline PSR J1903+327 \cite{36} & 1.667 $\pm$ 0.021 & 9.48 $\pm$ 0.03
\\
\hline PSR J1614-2230 \cite{37} & 1.97 $\pm$ 0.04 & 9.69 $\pm$ 0.2
\\
\hline Vela X-1 \cite{31} & 1.77 $\pm$ 0.08 & 9.56 $\pm$ 0.08
\\
\hline EXO 1785-248 \cite{38} & 1.30 $\pm$ 0.2 & 10.10 $\pm$ 0.44
\\
\hline
\end{tabular}
\end{center}
\end{table}
\begin{table}\caption{Values of output parameters.}
\begin{center}
\begin{tabular}{|c|c|c|c|c|c|c|c|}
\hline Compact stars & $a$ & $b$  & $c$
\\
\hline  4U 1538-52  & 0.173578 & 0.600865 & 27.3614
\\
\hline  SAX J1808.4-3658 & 0.144361 & 0.223964 & 40.8846
\\
\hline Her X - 1  & 0.565274 & 0.292825 & 34.7571
\\
\hline  LMC X-4  & 0.489691 & 0.524276 & 27.5939
\\
\hline  SMC X-4  & 0.21132 & 0.458942 & 29.641
\\
\hline 4 U 1820 - 30  & 0.267691 & 0.53818 & 28.4139
\\
\hline  Cen X-3  & 0.171037 & 0.305731 & 41.1245
\\
\hline  4U 1608-52  & 0.391925 & 0.250406 & 37.029
\\
\hline  PSR J1903+327  & 0.53744 & 0.378168 & 31.6204
\\
\hline  PSR J1614-2230  & 0.367443 & 0.61498 & 26.3748
\\
\hline  Vela X-1  & 0.172711 & 0.747927 & 25.6442
\\
\hline EXO 1785-248  & 0.111626 & 0.600865 & 27.3614
\\
\hline
\end{tabular}
\end{center}
\end{table}

\section{Analysis of Different Physical Aspects}

We analyze physical characteristics of various stellar objects in
this section. The following subsections discuss the graphical
behavior of different physical quantities which provide valuable
insights in the configuration of stellar structures.

\subsection{Evolution of Matter Contents}

The behavior of fluid variables, i.e., energy density, radial and
tangential pressures in self-gravitating objects is essential for
understanding their internal structure and behavior in
$f(\mathbb{Q},T)$ theory. The corresponding field equations are
\begin{eqnarray}\nonumber
\rho&=&\frac{(\frac{r}{c})^{\frac{-2b^{2}}{1+b}}}{3(1+b)
((b-2)b-1)r^{2}(1+\varsigma)(2\varsigma-1)}\bigg[-(1+b)
(\frac{r}{c})^{{2b^{2}}{1+b}}\sigma(10\varsigma
\\\nonumber
&+&(b-2)b(3+(5(b-2)b-17)\varsigma))-((b-2)b-1)(\frac{r}{c})^{4
-\frac{2}{1+b}}\sigma
\\\label{41}
&\times&b(9(1+\varsigma)+(15+50\varsigma+b((49+5b((b-2)b-7))
\varsigma-6)))\bigg],
\\\nonumber
P_{r}&=&\frac{(\frac{r}{c})^{\frac{-2b^{2}}{1+b}}}{3(1+b)
((b-2)b-1)r^{2}(1+\varsigma)(2\varsigma-1)}\bigg[(1+b)
(\frac{r}{c})^{{2b^{2}}{1+b}}\sigma(6-2\varsigma
\\\nonumber
&+&(b-2)b(5((b-2)b-1)\varsigma-3))+((b-2)b-1)(\frac{r}{c})
^{4-\frac{2}{1+b}}\sigma(9(1+\varsigma)
\\\label{42}
&+&b(21+38\varsigma+b(6+25\varsigma+b((5(b-2)b-23)\varsigma-6))))\bigg],
\\\nonumber
P_{t}&=&\frac{\sigma}{3r^{2}(\varsigma+2\varsigma^{2}-1)}
\bigg[\big(3-(b-2)b(6+(b-2)b(\varsigma-3)-\varsigma)
+4\varsigma\big)
\\\nonumber
&\times&\big((b-2)b-1\big)^{-1}+\big(r^{6}(\frac{r}{c})
^{-2b{-4}{1+b}}(3((b-2)b-1)(b^{3}-3-5b)
\\\label{43}
&+&(9+b(14+b(b+2b^{2}-b^{3}-5)))\varsigma)\big)\big
((1+b)c^{6}\big)^{-1}\bigg].
\end{eqnarray}
Figures \textbf{2} and \textbf{3} display graphical representations
of the fluid parameters and their derivatives for each star
candidate. The energy density and pressure components corresponding
to each star candidate exhibit monotonically decreasing behavior as
the radial distance increases. Additionally, the radial pressure
vanishes at the boundary of the star. Figure \textbf{3} demonstrates
that the derivative of fluid parameters is zero at the center and
becomes negative as one moves away from the core. This
characteristic confirms the highly compact configuration of the
proposed stars in the context of the $f(\mathbb{Q},T)$ theory.

Anisotropy $(\Delta=P_t-P_r)$ determines the directionally dependent
properties of a physical object \cite{26}. If anisotropy is
positive, the pressure is directed outward creating a repulsive
force $(\Delta>0)$. Conversely, negative anisotropy $(\Delta<0)$
manifests that the pressure is directed inward. Figure \textbf{4}
represents that 4 U 1820-30, Cen X-3, 4U 1608-52, PSR J1903+327, PSR
J1614-2230, Vela X-1 stars show positive anisotropy and 4U 1538-52,
Her X-1, LMC X-4, EXO 1785-248, SAX J1808.4-3658 compact stars
represent negative behavior. The anisotropy vanishes corresponding
to SMC X-4 compact star.
\begin{figure}
\epsfig{file=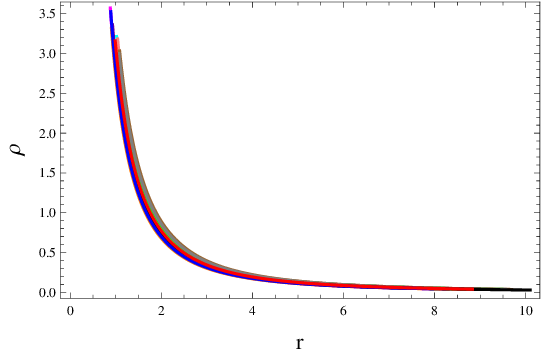,width=.5\linewidth}
\epsfig{file=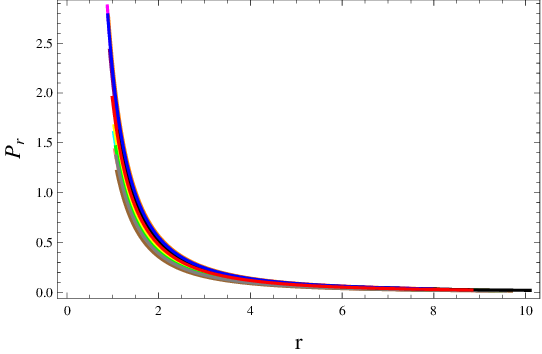,width=.5\linewidth}\center
\epsfig{file=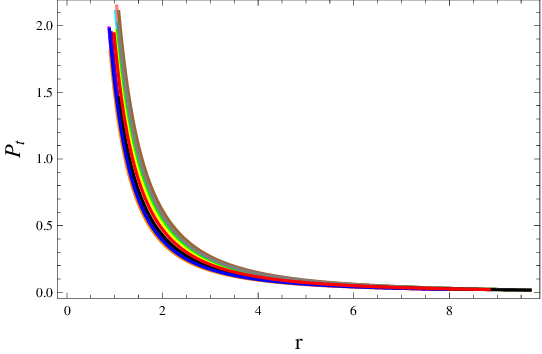,width=.5\linewidth}\caption{Behavior of fluid
parameters for $\sigma=1.5$ and $\varsigma=-0.4$.}
\end{figure}
\begin{figure}
\epsfig{file=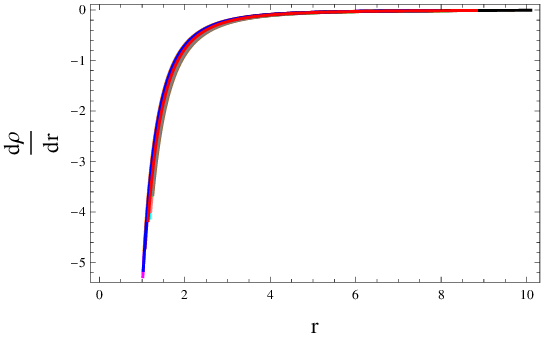,width=.5\linewidth}
\epsfig{file=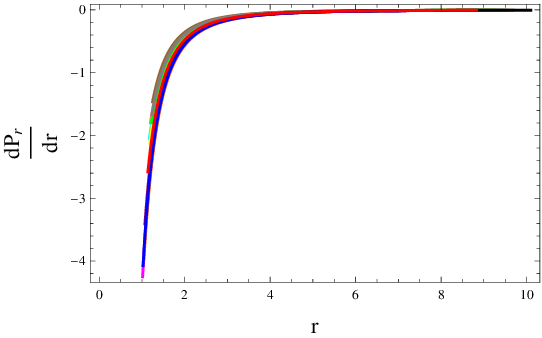,width=.5\linewidth}\center
\epsfig{file=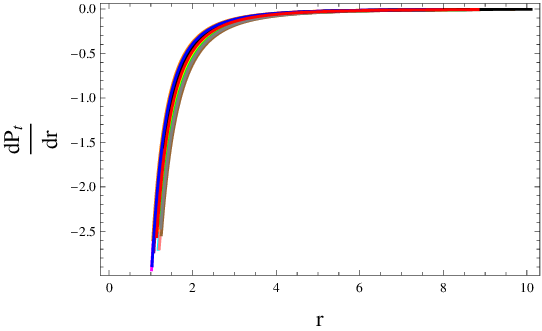,width=.5\linewidth}\caption{Graphs of gradient
of fluid parameters for $\sigma=1.5$ and $\varsigma=-0.4$.}
\end{figure}
\begin{figure}\center
\epsfig{file=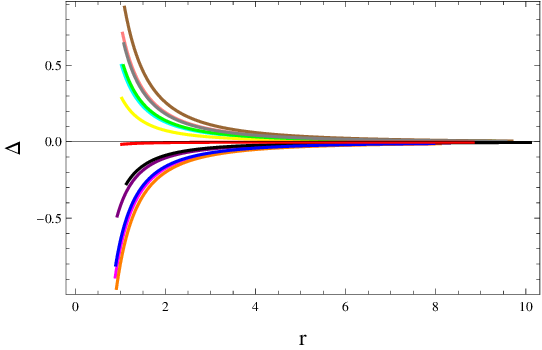,width=.5\linewidth}\caption{Behavior of
anisotropy for $\sigma=1.5$ and $\varsigma=-0.4$.}
\end{figure}

\subsection{Analysis of Energy Bounds}

These are the mathematical inequalities that place constraints on
the fluid parameters to describe the distribution of energy and
momentum in spacetime. By imposing these constraints, researchers
can explore the viability of various cosmic structures, classified
as null $(0\leq P_{r}+\rho,~0\leq P_{t}+\rho)$, dominant $(0\leq
\rho\pm P_{r},~ 0\leq \rho\pm P_{t})$,  weak $(0\leq P_{r}+\rho,~
0\leq P_{t}+\rho,~ 0\leq \rho)$ and strong $(0\leq P_{r}+\rho,~
0\leq P_{t}+\rho,~ 0\leq P_{r}+2P_{t}+\rho)$. energy constraints.
These energy bounds have a significant impact on the existence of
viable cosmic objects in spacetime. Figure \textbf{5} demonstrates
that the proposed star candidates are physically viable due to the
existence of normal matter inside the stars as all the energy
constraints are satisfied in the presence of $f(\mathbb{Q},T)$
terms.
\begin{figure}
\epsfig{file=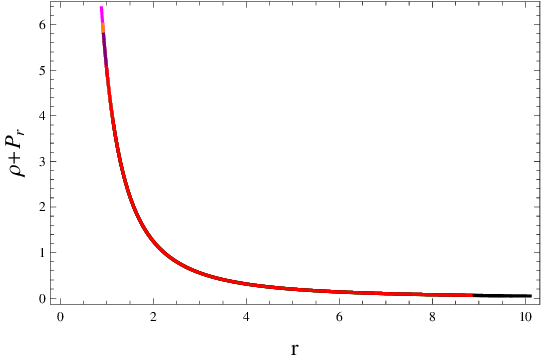,width=.5\linewidth}
\epsfig{file=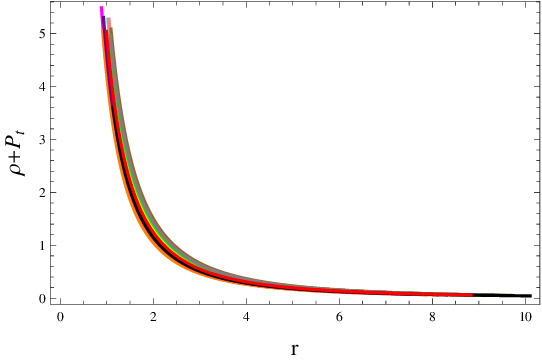,width=.5\linewidth}
\epsfig{file=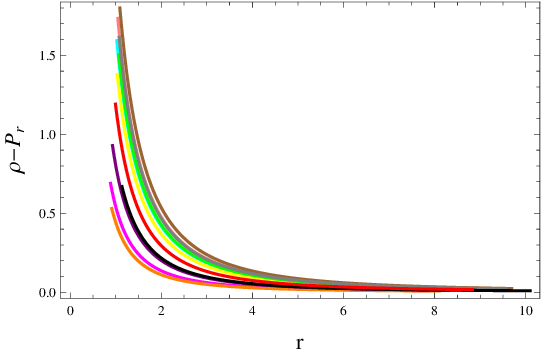,width=.5\linewidth}
\epsfig{file=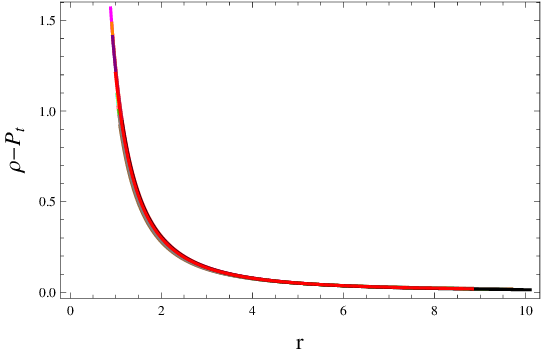,width=.5\linewidth}\caption{Behavior of energy
bounds versus $r$ for $\sigma=1.5$ and $\varsigma=-0.4$.}
\end{figure}

\subsection{State Parameters}

Equation of state parameters describe the relation between pressure
and energy density in various physical systems. For a physically
viable model, the radial $(\omega_{r}=\frac{P_{r}}{\rho})$ and
transverse $(\omega_{t}=\frac{P_{t}}{\rho})$ parameters must lie in
[0,1] \cite{27}. Using Eqs.(\ref{41})-(\ref{43}), we have
\begin{eqnarray}\nonumber
\omega_{r}&=&-\bigg[(1+b)(\frac{r}{c})^{\frac{2b^{2}}{1+b}}
(6-2\varsigma+(b-2)b(5((b-2)b-1)\varsigma-3))((b-2)
\\\nonumber
&\times&b-1)(\frac{r}{c})^{4-\frac{2}{1+b}}(9(1+\varsigma)
+b(21+38\varsigma+b(6+25\varsigma+b((5(b-2)b
\\\nonumber
&-&23)\varsigma-6))))\bigg]\bigg[(1+b)(\frac{r}{c})^{\frac{2b^{2}}
{1+b}}(10\varsigma+(b-2)b(3+(5(b-2)b-17)
\\\nonumber
&\times&\varsigma))+((b-2)b-1)(\frac{r}{c})^{4-\frac{2}{1+b}}
(9(1+\varsigma)+b(15+50\varsigma+b((49+5b
\\\nonumber
&\times&((b-2)b-7))\varsigma-6)))\bigg]^{-1},
\\\nonumber
\omega_{t}&=&\bigg[((1+b)(\frac{r}{c})^{{2b^{2}}{1+b}}+(1+b)
(-1+(b-2)b)(\frac{r}{c})^{4-\frac{2}{1+b}})(4c^{6}(\frac{r}{c})
^{2b+\frac{4}{1+b}}
\\\nonumber
&\times&(1+b)(-3+(b-2)b(6+(b-2)b(\varsigma-3)-\varsigma)
-4\varsigma)+4r^{6}(-3(-1
\\\nonumber
&\times&b(b-2))^{2}(b^{3}-3-5b)+(9+b(32+b(14+b(b(9+b(2+(b-4)
\\\nonumber
&\times&b))-23))))\varsigma))\bigg]\bigg[(1+b)(((b-2)b-1)r^{6}
+c^{6}(\frac{r}{c})^{2b+\frac{4}{1+b}})(4(1+b)
\\\nonumber
&\times&(\frac{r}{c})^{\frac{2b^{2}}{1+b}}(10\varsigma+(b-2)b
(3+(5(b-2)b-17)\varsigma))+4((b-2)b-1)
\\\nonumber
&\times&(\frac{r}{c})^{4-\frac{2}{1+b}}(9(1+\varsigma)+b(15+50
\varsigma+b((49+5b((b-2)b-7))\varsigma-6))))\bigg]^{-1}.
\end{eqnarray}
The graphical analysis of equation of state parameters is given in
Figure \textbf{6} which shows that $\omega_{r}$ and $\omega_{t}$
satisfy the required viability condition of the considered stars.
\begin{figure}
\epsfig{file=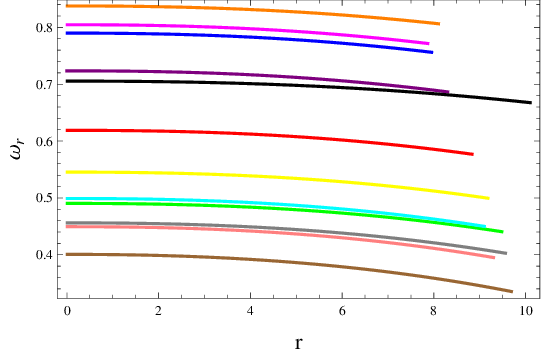,width=.5\linewidth}
\epsfig{file=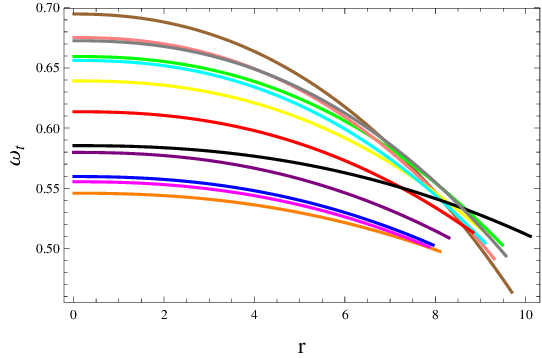,width=.5\linewidth}\caption{Behavior of
equation of state parameters versus $r$ for $\sigma=1.5$ and
$\varsigma=-0.4$.}
\end{figure}

\section{Stability Analysis}

Stability analysis examines the consequences of a small disturbance
on the structure of stars and whether it would go back to its
original state of equilibrium or undergo significant
transformations. Understanding the stability of cosmic structures is
paramount, as it provides valuable insights into their validity and
consistency. Stability analysis involves examining the conditions
that ensure cosmic structures remain stable against various
oscillation modes. The geometry of these structures and the
properties of the matter that forms and sustains them play crucial
roles in determining their stability. Here, we use the different
methods to analyze the stability of stars.

\subsection{Behavior of Various Forces}

The TOV equation determines the equilibrium structure of a static
spherical spacetime. This equation provides insights into how a
star's pressure and gravitational forces counterbalance to preserve
its equilibrium state. This significantly contributes to comprehend
compact star's internal structure and properties, playing a central
role in studying various astrophysical objects. This equation is
expressed as \cite{28}
\begin{equation}\label{50}
\frac{M_{G}(r)e^\frac{\lambda-\xi}{2}}{r^{2}}(\rho+P_{r})
+\frac{dp_{r}}{dr}-\frac{2}{r}(P_{t}-P_{r})=0,
\end{equation}
where
\begin{equation}\nonumber
M_{G}(r)=4\pi \int (T^{t}_{t}-T^{r}_{r} -
T^{\theta}_{\theta}-T^{\phi}_{\phi})r^{2}
e^{\frac{\lambda+\xi}{2}}dr.
\end{equation}
Solving this equation, we have
\begin{equation}\nonumber
M_{G}(r)=\frac{1}{2}r^{2}e^{\frac{\xi-\lambda}{2}}\lambda'.
\end{equation}
Equation (\ref{50}) turns out to be
\begin{equation}\nonumber
\frac{\lambda'(\rho+P_{r})}{2}+P_{r}'-\frac{2}{r}\Delta=0.
\end{equation}
This is the combination of different forces acting on the system
such as gravitational
$(\mathcal{F}_{g}=\frac{\lambda'(\rho+P_{r})}{2})$, hydrostatic
$(\mathcal{F}_{h}=\frac{dP_{r}}{dr})$ and anisotropic
$(\mathcal{F}_{a}=\frac{2}{r}\Delta)$ forces. Using
Eqs.(\ref{41})-(\ref{43}), we obtain
\begin{eqnarray}\nonumber
\mathcal{F}_{g}&=&2\sigma((b-2)b-1)\bigg[-1+\frac{b(1-(b-2)b)r^{4}
(\frac{r}{c})^{\frac{-2(1+b^{2})}{1+b}}}{(1+b)c^{4}}\bigg]
\bigg[r^{3}(1+\varsigma)\bigg]^{-1},
\\\nonumber
\mathcal{F}_{a}&=&2\sigma((b-1)b-1)\bigg[\frac{1+(b-3)b}{(b-2)b-1}
+\frac{b((b-1)b-4)r^{4}(\frac{r}{c})^{\frac{-2(1+b^{2})}
{1+b}}}{(1+b)c^{4}}\bigg]
\\\nonumber
&\times&\bigg[r^{3}(1+\varsigma)\bigg]^{-1},
\\\nonumber
\mathcal{F}_{h}&=&2(\frac{r}{c})^{\frac{-2b^{2}}{1+b}}\bigg[-(b+1)^{2}
(\frac{r}{c})^{\frac{2b^{2}}{1+b}}(6-2\beta+(b-2)b(-3+5(-1+(b-2)
\\\nonumber
&\times&b)\beta))-b(1+b-3b^{2}+b^{3})(\frac{r}{c})^{4-\frac{2}
{1+b}}(9(1+\beta)+b(21+38\beta+b(25\beta
\\\nonumber
&+&6+b(-6+(-23+5(b-2)b)\beta))))\bigg]\bigg[3(1+b)^{2}
(-1+(b-2)b)r^{3}
\\\nonumber
&\times&(1+\beta)(2\beta-1)\bigg]^{-1}.
\end{eqnarray}
Figure \textbf{7} shows that the our considered CSs are in
equilibrium state as the total effect of all forces is zero.
\begin{figure}\center
\epsfig{file=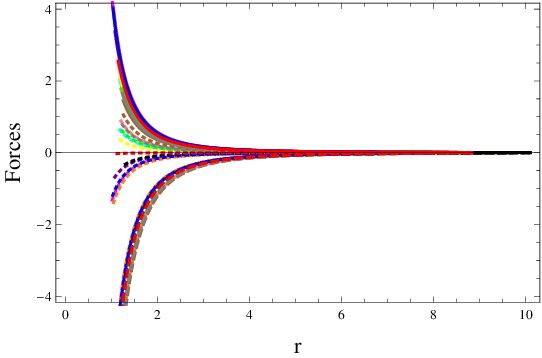,width=.5\linewidth}\caption{Graph of
Tolman-Oppenheimer-Volkoff equation for $\sigma=1.5$ and
$\varsigma=-0.4$.}
\end{figure}

\subsection{Sound Speed}

Sound speed is a fundamental property that characterizes how
pressure waves propagate through a medium. It is related to the
stiffness of the material inside the star and is crucial for
understanding its response to internal and external disturbances.
Causality condition can be used to examine the stability of stellar
objects, which assures that the no information cannot exceed the
speed of light. Accordingly, radial $(u_{r}=\frac{dP_{r}}{d\rho})$
and tangential $(u_{t} =\frac{dP_{t}}{d\rho})$ components of sound
speed should lie in the interval of $[0,1]$ \cite{29}. Sound speed's
components are
\begin{eqnarray}\nonumber
u_{r}&=&-\bigg[(1+b)^{2}(\frac{r}{c})^{{2b^{2}}{1+b}}(6-2\varsigma
+(b-2)b(-3+5(-1+(b-2)b)\varsigma))
\\\nonumber
&+&b(1+b-3b^{2}+b^{3})(\frac{r}{c})^{4-\frac{2}{1+b}}(9(1+\varsigma)
+b(21+38\varsigma+b(6+25\varsigma
\\\nonumber
&+&b((5(b-2)b-23)\varsigma-6))))\bigg]\bigg[(1+b)^{2}(\frac{r}{c})
^{{2b^{2}}{1+b}}(10\varsigma+(b-2)b(3
\\\nonumber
&+&(5(b-2)b-17)\varsigma))+b(1+b-3b^{2}+b^{3})(\frac{r}{c})^{4
-\frac{2}{1+b}}(9(1+\varsigma)
\\\nonumber
&+&b(15+50\varsigma+b((49+5b((b-2)b-7))\varsigma-6)))\bigg]^{-1},
\\\nonumber
u_{t}&=&\bigg[(1+b)^{2}(\frac{r}{c})^{{2b^{2}}{1+b}}(-3+(b-2)b
(6+(b-2)b(\varsigma-3)-\varsigma)-4\varsigma)
\\\nonumber
&+&(b(1+b-3b^{2}+b^{3})r^{4}(\frac{r}{c})^{\frac{-2}{1+b}}(-3(
(b-2)b-1)(-3-5b+b^{3})+
\\\nonumber
&\times&(-9+b(-14+b(5+b((b-2)b-1))))\varsigma))/c^{4})\bigg]
\bigg[(1+b)^{2}(\frac{r}{c})^{{2b^{2}}{1+b}}(10
\\\nonumber
&\times&\varsigma+(b-2)b(3+(-17+5(b-2)b)\varsigma))+b(1+b-3b^{2}
+b^{3})(\frac{r}{c})^{4-\frac{2}{1+b}}
\\\nonumber
&\times&(9(1+\varsigma)+b(15+50\varsigma+b((49+5b((b-2)b-7))
\varsigma-6)))\bigg]^{-1}.
\end{eqnarray}
Another method to study the stability of solutions is the  Herrera
cracking technique $(0\leq\mid u_{t}-u_{r}\mid\leq1)$ \cite{29a}.
The violation of this condition indicates that the compact stars are
in an unstable state, otherwise, it ensures a stable state. Figure
\textbf{8} shows that static spherical solutions are stable as they
fulfill the necessary constraints.
\begin{figure}
\epsfig{file=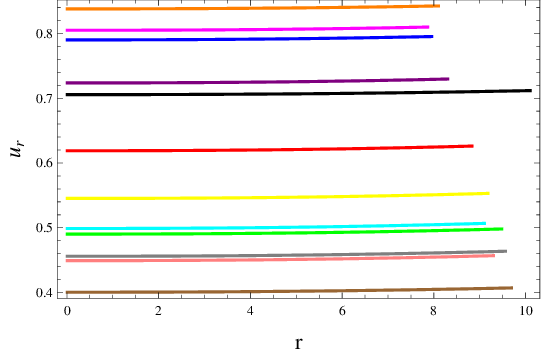,width=.5\linewidth}
\epsfig{file=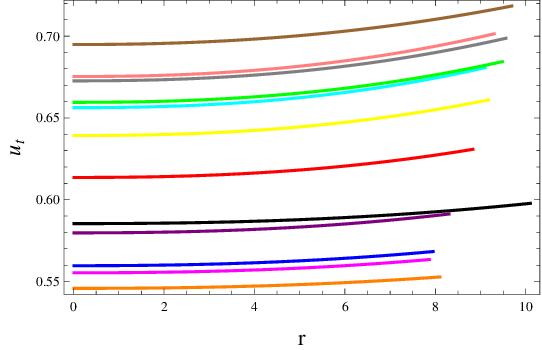,width=.5\linewidth}\center
\epsfig{file=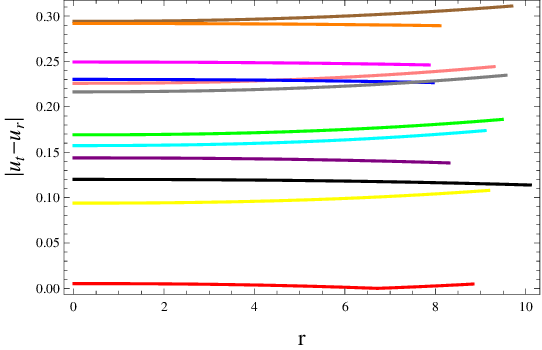,width=.5\linewidth}\caption{Behavior of sound
speed for $\sigma=1.5$ and $\varsigma=-0.4$.}
\end{figure}

\subsection{Adiabatic Index}

It characterizes the relationship between pressure and density
changes in the interior of stars. It is a key parameter in
astrophysics indicating how pressure responds to density variations
in a stellar system and is extensively used in the field of
astrophysics. Chandrasekhar \cite{30} developed a criteria for the
stability $(\Gamma>1.33)$ and behavior of astrophysical systems. It
is calculated as
\begin{eqnarray}\nonumber
\Gamma_{r}=\frac{\rho+P_{r}}{P_{r}} u_{r},\quad
\Gamma_{t}=\frac{\rho+P_{t}}{P_{t}} u_{t}.
\end{eqnarray}
The components of adiabatic index are given by
\begin{eqnarray}\nonumber
\Gamma_{r}&=&\bigg[-6((b-2)b-1)((1+b)(\frac{r}{c})^{\frac{2b^{2}}
{1+b}}+b((b-2)b-1)(\frac{r}{c})^{4-\frac{2}{1+b}})(2\varsigma-1)
\\\nonumber
&\times&((1+b)^{2}(\frac{r}{c})^{\frac{2b^{2}}{1+b}}(6-2\varsigma
+(b-2)b(5((b-2)b-1)\varsigma-3))+b(1+b
\\\nonumber
&-&3b^{2}+b^{3})(\frac{r}{c})^{4-\frac{2}{1+b}}(9(1+\varsigma)
+b(21+38\varsigma+b(6+25\varsigma+b((5(b-2)b
\\\nonumber
&-&23)\varsigma-6)))))\bigg]\bigg[((1+b)^{2}(\frac{r}{c})^{\frac
{2b^{2}}{1+b}}(10\varsigma+(b-2)b(3+(5(b-2)b-17)
\\\nonumber
&\times&\varsigma))+b(1+b-3b^{2}+b^{3})(\frac{r}{c})^{4-\frac{2}
{1+b}}(9(1+\varsigma)+b(50\varsigma+15+b((49+5
\\\nonumber
&\times&b((b-2)b-7))\varsigma-6))))((1+b)(\frac{r}{c})^{\frac
{2b^{2}}{1+b}}(6-2\varsigma+b(b-2)(5((b-2)
\\\nonumber
&\times&b-1)\varsigma-3))+((b-2)b-1)(\frac{r}{c})^{4-\frac{2}
{1+b}}(9(1+\varsigma)+b(21+38\varsigma
\\\nonumber
&+&+b(6+25\varsigma+b((5(b-2)b-23)\varsigma-6)))))\bigg]^{-1},
\\\nonumber
\Gamma_{t}&=&\bigg[3(\frac{r}{c})^{\frac{-2}{1+b}}((1+b)(1+(b-3)
(b-2)b(1+b))(\frac{r}{c})^{\frac{2b^{2}}{1+b}}+(b-2)b
\\\nonumber
&\times&((b-2)b-1)(b^{3}-6b-3)(\frac{r}{c})^{4-\frac{2}{1+b}})
(2\varsigma-1)((1+b)^{2}c^{4}(\frac{r}{c})^{\frac{2(1+b^{2})}
{1+b}}(-3
\\\nonumber
&+&(b-2)b(6+(b-2)b(\varsigma-3)-\varsigma)-4\varsigma)+b(1+b
-3b^{2}+b^{3})r^{4}(-3
\\\nonumber
&\times&((b-2)b-1)(b^{3}-3-5b)+(b(b(5+b((b-2)b-1))-14)-9)
\varsigma))\bigg]
\\\nonumber
&\times&\bigg[c^{4}((1+b)(\frac{r}{c})^{\frac{2b^{2}}{1+b}}
(-3+(b-2)b(6+(b-2)b(\varsigma-3)-\varsigma)-4\varsigma)((b
\\\nonumber
&-&2)b-1)(\frac{r}{c})^{4-\frac{2}{1+b}}(-3(b(b-2)-1)(b^{3}-3-5b)
+(b(b(5+b((b-2)
\\\nonumber
&\times&b-1))-14)-9)\varsigma))((1+b)^{2}(\frac{r}{c})^{\frac
{2b^{2}}{1+b}}(10\varsigma+(b-2)b(3+(5(b-2)b
\\\nonumber
&-&17)\varsigma))+b(1+b-3b^{2}+b^{3})(\frac{r}{c})^{4-\frac{2}
{1+b}}(9(1+\varsigma)+b(15+50\varsigma+b((49
\\\nonumber
&+&5b((b-2)b-7))\varsigma-6)))) \bigg]^{-1}.
\end{eqnarray}
Figure \textbf{9} shows that our system is stable in the presence of
correction terms as it satisfies the required limit. Hence, we
obtain viable and stable compact stars in $f(\mathbb{Q},T)$ theory.
\begin{figure}
\epsfig{file=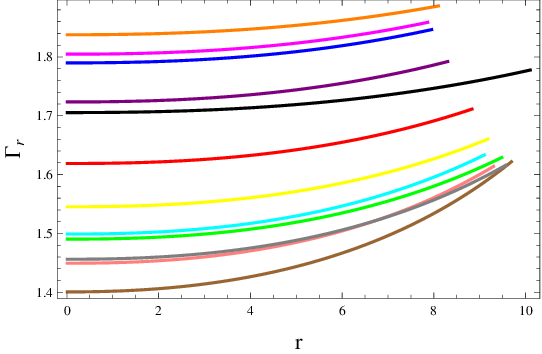,width=.5\linewidth}
\epsfig{file=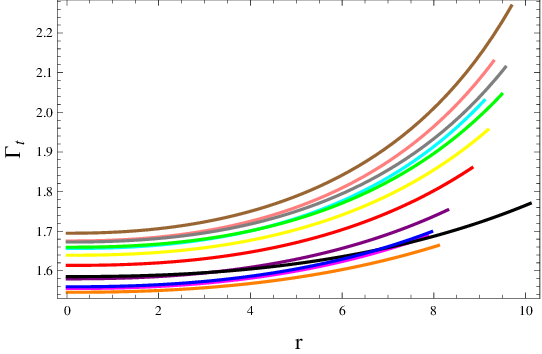,width=.5\linewidth}\caption{Graphs of adiabatic
index for $\sigma$=1.5 and $\varsigma$=-0.4.}
\end{figure}

\section{Final Remarks}

The exploration of compact stars has been an intriguing field of
study in theoretical physics over the last twenty years. This
research is centered on investigating compact stellar structures in
the framework of modified $f(\mathbb{Q}, T)$ theory. The aim is to
acquire profound insights into cosmic mysteries. This modified
theory provides a promising avenue for elucidating phenomena
associated with the dark universe. This theory is a captivating
method that does not include any exotic matter component. This
non-conservative theory investigates the effect of non-geodesic
motion on the particles. By investigating the behavior of compact
stars in this modified theory, we gain insights into gravitational
interactions on both galactic and cosmological scales. This sheds
light on the nature of these stellar components and their influence
on stellar structures. The gravitational conditions in compact
stellar objects reach their most extreme limits, making them a
crucial context for studying gravity behavior under a high curvature
regime. This exploration provides valuable information on the
characteristics of compact stars, enhancing our overall
understanding of the fundamental interactions that shape the
universe. This research significantly advances our comprehension of
gravity, opening new avenues for a deeper understanding of the
cosmos and its governing forces.

We have examined the viability and stability of compact stars in
this theory. Furthermore, the values of $a,b,c$ are determined by
smooth matching of the interior (static spherical) and the exterior
(Schwarzschild) spacetimes, and their values are listed in Table
\textbf{3}. We have assumed specific model of this modified gravity
to examine viable and stable compact stars through different
parameters. The main results are given as follows.

The metric elements are consistent and meet necessary conditions
with a minimum value at the center (Figure \textbf{1}). The behavior
of fluid parameters is positive and regular in the interior of the
proposed stars and decreases at the boundary (Figure \textbf{2}).
The derivative of fluid parameters indicates a dense distribution in
the compact stars (Figure \textbf{3}). The anisotropic pressure in
the stars is directed outward for 4 U 1820-30, Cen X-3, 4U 1608-52,
PSR J1903+327, PSR J1614-2230, Vela X-1 stars, whereas pressure is
inward for 4U 1538-52, Her X-1, LMC X-4, EXO 1785-248, SAX
J1808.4-3658 compact stars. Also, the anisotropy vanishes
corresponding to SMC X-4 compact star (Figure \textbf{4}). All
energy bounds are satisfied confirming the presence of normal matter
in the interior of the compact stars (Figure \textbf{5}). The range
of equation of state parameters lies between 0 and 1, indicating the
viability of the considered model (Figure \textbf{6}).
Tolman-Oppenheimer-Volkoff equation suggests that gravitational,
hydrostatic and anisotropic forces have a null impact with respect
to proposed compact stellar objects, indicating that they are in
equilibrium (Figure \textbf{7}). Stability conditions such as
causality conditions ($u_{r}$,$u_{t}\in[0,1]$), Herrera cracking
($0<|u_{t}- u_{r}|<1$) and the adiabatic index ($\Gamma>4/3$) are
satisfied, ensuring the stability of the compact stars under
considerations (Figures \textbf{8}-\textbf{9}).

We have examined whether considered compact stars maintain their
stability in this theory. Our comprehensive analysis of the obtained
solutions has yielded a dense profile for these compact stars. In
our analysis and investigation of the physical aspects, we have
obtained a more dense profile of compact stars. We have studied the
behavior of essential physical parameters, including metric
potentials, effective matter variables, the EoS parameters, redshift
function, energy conditions, TOV equation, sound speed and adiabatic
index, which characterize the stellar system. It is noteworthy to
emphasize that all the aforementioned physical parameters meet the
necessary conditions, underscoring the presence of viable and stable
compact stars in this modified framework. The chosen factors for
analyzing the feasibility and stability of the solution have
satisfied their specified limits.

Notably, we observed that all parameters reach their maximum values
when compared to general relativity \cite{38b}-\cite{38bb} and other
modified gravitational theories \cite{38bbb}. In the realm of
$f(\textrm{R})$ theory, the results indicate the instability of the
Her X-1 compact star associated with the second gravity model due to
the limited range satisfied by the physical quantities \cite{38bbb}.
Furthermore, in the framework of $f(\textrm{R},\textrm{T}^{2})$
theory, it is found that compact stars are neither physically viable
nor stable at the center \cite{38bbbb}. In light of these findings,
it can be concluded that all considered compact stars exhibit both
physical viability and stability in this modified theory.
Consequently, our results suggest that viable and stable compact
stars can exist in this modified theory. Therefore, we conclude that
the solutions we have obtained are physically valid, providing
stable and viable structures for anisotropic compact objects.

\vspace{0.25cm}

\section*{Appendix A: Non-Metricity Scalar}
\renewcommand{\theequation}{A\arabic{equation}}
\setcounter{equation}{0}

According to Eqs.(\ref{25}) and (\ref{27}), we have
\begin{eqnarray}\nonumber
\mathbb{Q} &\equiv& -\mathrm{g}^{iv}
(\mathbb{L}^{k}_{~ui}\mathbb{L}^{u}_{~vk} -
\mathbb{L}^{k}_{~uk}\mathbb{L}^{u}_{~iv}),
\\\nonumber
\mathbb{L}^{k}_{~ui}&=&-\frac{1}{2} \mathrm{g}^{kj}(\mathbb{Q}_{iuj}
+\mathbb{Q}_{uji}-\mathbb{Q}_{jiu}),
\\\nonumber
\mathbb{L}^{u}_{~vk}&=&-\frac{1}{2} \mathrm{g}^{uj}(\mathbb{Q}_{kvj}
+\mathbb{Q}_{vjk}-\mathbb{Q}_{jkv}),
\\\nonumber
\mathbb{L}^{k}_{~ui} &=& -\frac{1}{2}
\mathrm{g}^{kj}(\mathbb{Q}_{kuj}
+\mathbb{Q}_{ujk}-\mathbb{Q}_{jku}),
\\\nonumber
&=&-\frac{1}{2}(\bar{\mathbb{Q}}_{u}
+\mathbb{Q}_{u}-\bar{\mathbb{Q}}_{u})=-\frac{1}{2} \mathbb{Q}_{u},
\\\nonumber
\mathbb{L}^{u}_{~iv}&=& -\frac{1}{2}\mathrm{g}^{uj}(\mathbb{Q}_{vij}
+\mathbb{Q}_{ijv}-\mathbb{Q}_{jvi}).
\end{eqnarray}
Thus, we have
\begin{eqnarray}\nonumber
-\mathrm{g}^{iv}\mathbb{L}^{k}_{~ui} \mathbb{L}^{u}_{~vk}&=&
-\frac{1}{4}\mathrm{g}^{iv}\mathrm{g}^{kj} \mathrm{g}^{uj}
(\mathbb{Q}_{iuj}+\mathbb{Q}_{uji} -\mathbb{Q}_{jiu})
\\\nonumber
&\times&(\mathbb{Q}_{kvj}+\mathbb{Q}_{vjk} -\mathbb{Q}_{jkv}),
\\\nonumber
&=&-\frac{1}{4}(\mathbb{Q}^{vjk}+\mathbb{Q}^{jkv} -\mathbb{Q}^{kvj})
\\\nonumber
&\times&(\mathbb{Q}_{kvj}+\mathbb{Q}_{vjk} -\mathbb{Q}_{jkv}),
\\\nonumber
&=&-\frac{1}{4}(2\mathbb{Q}^{vjk}\mathbb{Q}_{jkv} -
\mathbb{Q}^{vjk}\mathbb{Q}_{vjk}),
\\\nonumber
\mathrm{g}^{iv}\mathbb{L}^{k}_{~uk}\mathbb{L}^{u}_{~iv}&=&
\frac{1}{4}\mathrm{g}^{iv}\mathrm{g}^{uj}\mathbb{Q}_{j}
(\mathbb{Q}_{vij}+\mathbb{Q}_{ijv} -\mathbb{Q}_{jvi}),
\\\nonumber
&=&\frac{1}{4}\mathbb{Q}^{j}(2\bar{\mathbb{Q}_{j}}- \mathbb{Q}_{j}),
\\\nonumber
\mathbb{Q}&=& -\frac{1}{4}(\mathbb{Q}^{kvj}\mathbb{Q}_{kvj}
+2\mathbb{Q}^{kvjk}\mathbb{Q}_{jkv}
\\\nonumber
&-&2\mathbb{Q}^{j}\bar{\mathbb{Q}_{j}}+\mathbb{Q}^{j}\mathbb{Q}_{j}).
\end{eqnarray}
According to Eq.(\ref{25}), we obtain
\begin{eqnarray}\nonumber
\mathbb{P}^{kiv}&=&\frac{1}{4}[-\mathbb{Q}^{kiv}
+\mathbb{Q}^{ikv}+\mathbb{Q}^{vki} +\mathbb{Q}^{k}\mathrm{g}^{iv}
\\\nonumber
&-&\bar{\mathbb{Q}^{k}}\mathrm{g}^{iv}-\frac{1}{2} (\mathrm{g}^{ki}
\mathbb{Q}^{v}+\mathrm{g}^{kv}\mathbb{Q}^{i})],
\\\nonumber
-\mathbb{Q}_{kiv}\mathbb{P}^{kiv} &=&
-\frac{1}{4}[-\mathbb{Q}_{kiv}\mathbb{Q}^{kiv}
\\\nonumber
&+&\mathbb{Q}_{kiv}\mathbb{Q}^{ikv} +\mathbb{Q}^{vki}
\mathbb{Q}_{kiv}+\mathbb{Q}_{kiv} \mathbb{Q}^{k}\mathrm{g}^{iv}
\\\nonumber
&-&\mathbb{Q}_{kiv}\bar{\mathbb{Q}^{k}} \mathrm{g}^{iv}
-\frac{1}{2}\mathbb{Q}_{kiv}(\mathrm{g}^{ki} \mathbb{Q}^{v}
+\mathrm{g}^{kv}\mathbb{Q}^{i})],
\\\nonumber
&=& -\frac{1}{4}(-\mathbb{Q}_{kiv}\mathbb{Q}^{kiv}
+2\mathbb{Q}_{kiv}\mathbb{Q}^{ikv}+\mathbb{Q}^{k}
\mathbb{Q}_{k}-2\tilde{\mathbb{Q}^{k}}\mathbb{Q}_{k}),
\\\nonumber
&=& \mathbb{Q}.
\end{eqnarray}

\section*{Appendix B: Variation of Non-Metricity Scalar}
\renewcommand{\theequation}{B\arabic{equation}}
\setcounter{equation}{0}

All the non-metricity tensors are given as
\begin{eqnarray}\nonumber
\mathbb{Q}_{kiv}&=&\nabla_{k}\mathrm{g}_{iv},
\\\nonumber
\mathbb{Q}^{k}~_{iv}&=&\mathrm{g}^{ku} \mathbb{Q}_{uiv}
=\mathrm{g}^{ku}\nabla_{u}\mathrm{g}_{iv}
=\nabla^{k}\mathrm{g}_{iv},
\\\nonumber
\mathbb{Q}_{k~~v}^{~~i}&=&\mathrm{g}^{ij}\mathbb{Q}_{kjv}
=\mathrm{g}^{ij}\nabla_{k}\mathrm{g}_{jv}
=-\mathrm{g}_{ij}\nabla_{k}\mathrm{g}^{ij},
\\\nonumber
\mathbb{Q}_{ki}^{~~v} &=& \mathrm{g}^{vj}\mathbb{Q}_{kij}
=\mathrm{g}^{vj}\nabla_{k}\mathrm{g}_{ij}
=-\mathrm{g}_{ij}\nabla_{k}\mathrm{g}^{vj},
\\\nonumber
\mathbb{Q}^{ki}_{~~v}&=&
\mathrm{g}^{ij}\mathrm{g}^{ku}\nabla_{u}\mathrm{g} _{jv}
=\mathrm{g}^{ij}\nabla^{k}\mathrm{g}_{vj}
=-\mathrm{g}_{jv}\nabla^{k}\mathrm{g}^{ij},
\\\nonumber
\mathbb{Q}^ {k~~v} _{~i} &=&
\mathrm{g}^{vj}\mathrm{g}^{ku}\nabla_{u}\mathrm{g} _{ij}
=\mathrm{g}^{vj}\nabla^{k}\mathrm{g}_{ij}
=-\mathrm{g}_{ij}\nabla^{k}\mathrm{g}^{vj},
\\\nonumber
\mathbb{Q}_{k}^{~~iv}&=&
\mathrm{g}^{ij}\mathrm{g}^{vu}\nabla_{k}\mathrm{g} _{ju}
=-\mathrm{g}^{ij}\mathrm{g}_{ju}\nabla_{k} \mathrm{g}^{vj}
=-\nabla_{k}\mathrm{g}^{iv}.
\end{eqnarray}
By using Eq.(\ref{26}), we have
\begin{eqnarray}\nonumber
\delta \mathbb{Q} &=&-\frac{1}{4} \delta(-\mathbb{Q}^{kvj}
\mathbb{Q}_{kvj}+2\mathbb{Q}^{kvj} \mathbb{Q}_{jkv}-2\mathbb{Q}^{j}
\bar{\mathbb{Q}_{j}}+\mathbb{Q}^{j}\mathbb{Q}_{j}),
\\\nonumber
&=&-\frac{1}{4}(-\delta \mathbb{Q}^{kvj} \mathbb{Q}_{kvj} -
\mathbb{Q}^{kvj}\delta \mathbb{Q}_{kvj} + 2\delta
\mathbb{Q}_{kvj}\mathbb{Q}^{jkv}
\\\nonumber
&+& 2 \mathbb{Q}^{kvj}\delta \mathbb{Q}_{jkv}-2\delta
\mathbb{Q}^{j}\bar{\mathbb{Q}_{j}}+\delta
\mathbb{Q}^{j}\mathbb{Q}_{j}-2 \mathbb{Q}^{j}\delta \bar
{\mathbb{Q}_{j}} + \mathbb{Q}^{j}\delta \mathbb{Q}_{j}),
\\\nonumber
&=&-\frac{1}{4}[\mathbb{Q}_{kvj}\nabla ^{k}\delta
\mathrm{g}^{vj}-\mathbb{Q}^{kvj} \nabla_{k}\delta
\mathrm{g}_{vj}-2\mathbb{Q}_{jkv} \nabla^{k}\delta \mathrm{g}^{vj}
\\\nonumber
&+&2\mathbb{Q}^{kvj}\nabla_{j}\delta \mathrm{g}_{kv}+
2\bar{\mathbb{Q}_{j}}\mathrm{g}^{iv}\nabla ^{j}\delta
\mathrm{g}_{iv}+2\mathbb{Q}^{j}\nabla^{u}\delta \mathrm{g}_{ju}
\\\nonumber
&+&2\bar{\mathbb{Q}_{j}} \mathrm{g}_{iv}\nabla^{j}\delta
\mathrm{g}^{iv}-\mathbb{Q}_{j}\nabla^{u}\mathrm{g} ^{iv}\delta
\mathrm{g}_{iv}-\mathbb{Q}_{j}\mathrm{g}_{iv}\nabla^{j}\delta
\mathrm{g}^{iv}
\\\nonumber
&-&\mathbb{Q}_{j} \mathrm{g}^{iv} \nabla_{j} \delta \mathrm{g}_{iv}
-\mathbb{Q}^{j}\mathrm{g}_{iv}\nabla_{j}\delta \mathrm{g}_{iv}].
\end{eqnarray}
We use the following relations to simplify the above equation
\begin{eqnarray}\nonumber
\delta \mathrm{g}_{iv}&=&-\mathrm{g}_{ik} \delta
\mathrm{g}^{ku}\mathrm{g}_{uv}-\mathbb{Q}^ {k v j} \nabla_{k}\delta
\mathrm{g}_{vj},
\\\nonumber
&=&-\mathbb{Q}^{kvj} \nabla_{k}(-\mathrm{g}_{vi}\delta
\mathrm{g}^{iu}\mathrm{g}_{uj}),
\\\nonumber
&=&2\mathbb{Q}_{~~j}^{kv}\mathbb{Q}_{kvi}\delta \mathrm{g}^{ij} +
\mathbb{Q}_{kuj}\nabla^{k}\mathrm{g}^{ij}
\\\nonumber
&=&2\mathbb{Q}_{~~v}^{ku}\mathbb{Q}_{kuv}\delta
\mathrm{g}^{iv}+\mathbb{Q}_{kvj}\nabla^{k} \mathrm{g}^{vj},
\\\nonumber
2\mathbb{Q}^{kvj}\nabla_{j}\delta \mathrm{g}_{kv}&=&
-4\mathbb{Q}_{i}^{~uj}\mathbb{Q}_{juv} \delta \mathrm{g}^{iv}-2
\mathbb{Q}_{vjk}\nabla^{k}\delta \mathrm{g}^{vj},
\\\nonumber
-2\mathbb{Q}^{j} \nabla^{u} \delta \mathrm{g}_{ju}&=&2\mathbb{Q}^{k}
\mathbb{Q}_{vki} \delta \mathrm{g}^{iv}+
2\mathbb{Q}_{i}\bar{\mathbb{Q}_{v}} \delta \mathrm{g}^{iv}
\\\nonumber
&+&2\mathbb{Q}_{v}\mathrm{g}_{kj}\nabla^{k} \mathrm{g}^{vj}.
\end{eqnarray}
Thus, we have
\begin{equation}\nonumber
\delta\mathbb{Q}=2\mathbb{P}_{kvj}\nabla^{k}\delta
\mathrm{g}^{vj}-(~\mathbb{P}_{iku}\mathbb{Q}_{v} ^{~ku}-2
\mathbb{P}_{kuv}\mathbb{Q}^{ku} _{~v})\delta \mathrm{g}^{iv},
\end{equation}
where
\begin{eqnarray}\nonumber
2\mathbb{P}_{kvj}&=&-\frac{1}{4}[2\mathbb{Q}
_{kvj}-2\mathbb{Q}_{jkv} -2\mathbb{Q}_{vjk}
\\\nonumber
&+&2(\bar{\mathbb{Q}}_{k}-\mathbb{Q}_{k})
\mathrm{g}_{vj}+2\mathbb{Q}_{v}\mathrm{g}_{ku}],
\\\nonumber
4(\mathbb{P}_{iku}\mathbb{Q}_{v}^{~~ku}-2
\mathbb{P}_{kuv}\mathbb{Q}^{ku} _{~~v})&=&2\mathbb{Q}^{ku}
_{~~v}\mathbb{Q}_{kui}-4\mathbb{Q}_{i}
^{~~ku}\mathbb{Q}_{ukv}+2\mathbb{Q}_{kiv} \bar{\mathbb{Q}}^{k}
\\\nonumber
&-&\mathbb{Q}^{k}\mathbb{Q}_{kiv}
+2\mathbb{Q}^{k}\mathbb{Q}_{vki}+2\mathbb{Q}
_{i}\bar{\mathbb{Q}_{v}}.
\end{eqnarray}\\
\textbf{Data Availability:} No data was used for the research
described in this paper.

\end{document}